\documentclass[twocolumn,floatfix,superscriptaddress,nofootinbib,amssymb,amsmath,aps,prc,eqsecnum]{revtex4-1}
\usepackage[dvips]{graphicx}
\usepackage{latexsym,epsfig,bm,psfrag,subfigure}
\usepackage{color}         
\usepackage[bookmarksnumbered,bookmarksopen,colorlinks,citecolor=blue,linkcolor=blue]{hyperref} %
\newcommand{\Tdagger}[2]{T^{#1}_{#2}\,}
\newcommand{\T}[2]{T^{#1}_{\,\,#2}\,}

\newcommand{\fig}[1]{Fig.~\ref{#1}}

\newcommand{\mpm}{M_{\ms +-}}
\newcommand{\tpm}{\theta_{\ms +-}}
\renewcommand{\vec}{\boldsymbol}
\newcommand{\usl}[1]{U_{\ms 1#1}}
\newcommand{\uslt}[1]{U_{\ms 2#1}}
\newcommand{\mr}{M}

\newcommand{\ssq}{\mathcal{S}_{\ms +-}^2}
\newcommand{\ep}{$e^+$-$e^-$}
\newcommand{\lis}{{}^{7}\mathrm{Li}}
\newcommand{\bee}{{}^{8}\mathrm{Be}}
\newcommand{\mc}{M_{\mathrm{c}}}
\newcommand{\mn}{M_{\mathrm{n}}}
\newcommand{\mnc}{M_{\mathrm{nc}}} 
\newcommand{\fdu}[2]{{#1}^{\dagger #2}}

\newcommand{\fd}[2]{{#1}_{#2}}
\newcommand{\ms}{\scriptscriptstyle}
\newcommand{\h}[2] {h_{\scriptscriptstyle #1 ^#2\hspace{-0.8mm}P_1}  }
\newcommand{\dm}[1]{d_{\ms M1\left(#1\right)}}
\newcommand{\deone}{d_{\ms E1}}
\newcommand{\detwo}[1]{d_{\ms E2,#1} }

\begin{document}

%\preprint{}


\title{Can nuclear physics explain the anomaly observed in the internal pair production in the Beryllium-8 nucleus?}


\author{Xilin Zhang}
\email{xilinz@uw.edu}
%%\thanks{}
\affiliation{Department of Physics, University of Washington, Seattle, WA \ \ 98195, USA}

\author{Gerald A. Miller}
\email{miller@phys.washington.edu}
%%\thanks{}
\affiliation{Department of Physics, University of Washington, Seattle, WA \ \ 98195, USA}

\date{\today}

\begin{abstract}
Recently the experimentalists in [PRL {\bf 116}, 042501 (2016)] announced observing an unexpected enhancement of the \ep pair production signal in one of the $\bee$ nuclear transitions. The following studies have been focused on possible explanations based on introducing new types of particle. In this work, we improve the nuclear physics modeling of the reaction by studying the pair emission anisotropy and the interferences between different multipoles in an effective field theory inspired framework, and examine their possible relevance to the anomaly. The connection between the previously measured on-shell photon production and the pair production in the same nuclear transitions is established. These improvements, absent in the original experimental analysis, should be included in extracting new particle's properties from the  experiment of this type. We then study the possibility of using the nuclear transition form factor to explain the anomaly. The reduction of the anomaly's significance by simply rescaling our predicted event count is also investigated.     
\end{abstract}

%\keywords{}


\maketitle

\section{Introduction}
It was announced in Ref.~\cite{Krasznahorkay:2015iga} that in the measurement of the \ep pair production in the $\bee$'s nuclear transition between one of its $1^+$ resonance and its ground state (GS, a narrow resonance), an unexpected enhancement of the signal was observed in the large \ep invariant mass region (about 17 MeV) and in the large pair correlation angle (near $140^\circ$) region. The observation has generated strong interest  in the particle physics community, because the anomaly could be explained by new types of particles (e.g.,~\cite{Krasznahorkay:2015iga,Feng:2016jff}). However, the nuclear physics model from Ref.~\cite{Rose:1949} as used by the experimentalists for simulating the pair  production \cite{Gulyas:2015mia} through virtual photon decay is incomplete. In the experiment, the initial state is a beam-target plane wave and  sets up a particular direction in the reaction, leading to anisotropy in the pair emission. Moreover, in the anomalous reaction channel, the E1 and M1 multipoles have similar weights and their interference is substantial. Furthermore, the on-shell photon production measurements \cite{Tilley:2004zz,Barker:1996,Zahnow1995,Schlueter1964,Mainsbridge1960} provide important constraints on the multipoles in the pair production. In this work, we set up a model inspired by the so-called Halo effective field theory (EFT) framework \cite{Hammer:2017tjm, Zhang:2015ajn}, taking into account the aforementioned factors which have not been addressed before \cite{Rose:1949}, calibrate it to the photon production data, and predict the pair production cross section. The results, as well as the approach, could be used for analyzing future experiment of this type.  Although a direct comparison to the current \ep data is not feasible due to the missing public information  about the experimental detector efficiency \cite{Gulyas:2015mia}, the shape comparisons are still valuable. We find that the model improvements are not able to explain the anomaly. We also evade the photon production constraint by invoking a hypothetical form factor for the M1 transition, and show that the form factor needed to explain the anomaly suggests an unrealistic large length scale on the order of $10$s fm for the $\bee$ nucleus. We then study how the anomaly's significance is modified when the normalizations of our event estimation are allowed to vary. In the following, section~\ref{sec:model} discusses the kinematics and our model; section~\ref{sec:calibration} is about the model calibration. We then present our pair-production results in section~\ref{sec:epresults}, and explore possible M1 transition form factor in section~\ref{sec:ff}. A short summary is provided in the end.

\section{Kinematics and the EFT-inspired model} \label{sec:model}

\begin{figure}
	\centering
		\includegraphics[width=6.5cm]{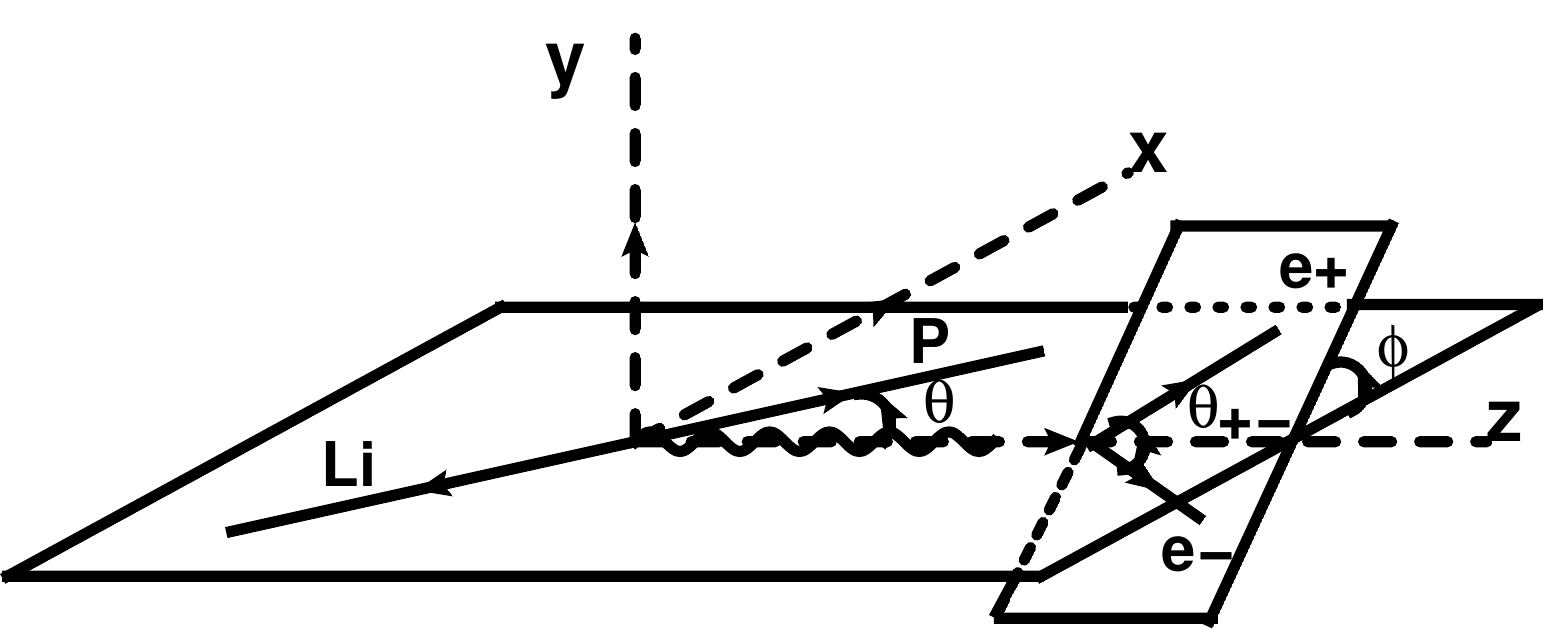}
		  \includegraphics[width=5cm]{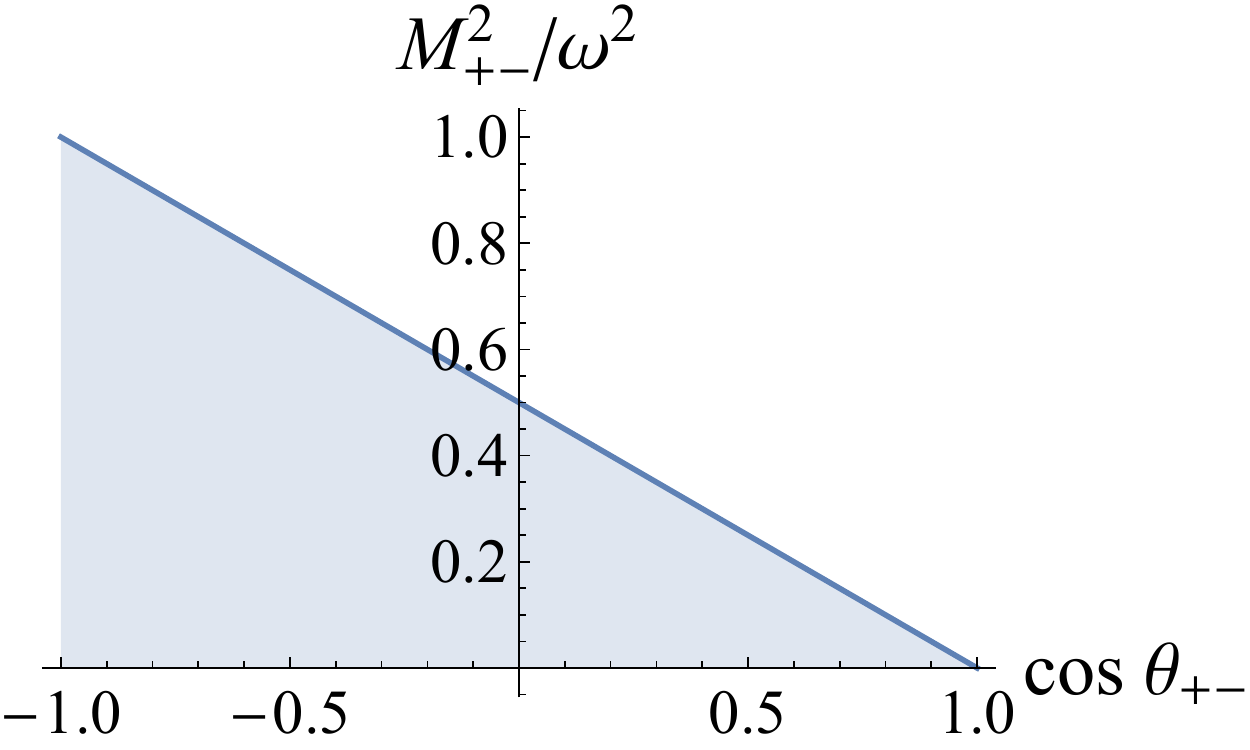} 
     \caption{The top shows the kinematics for the \ep pair production as well as the photon production (without the lepton line). The bottom plots the allowed phase space (shaded area) in terms of $\mpm$ and $\cos\tpm$ assuming $m_e=0$.} \label{fig:kinematics}
\end{figure} 

\fig{fig:kinematics} illustrates the relevant kinematic variables for both pair and photon productions in the proton--$\lis$ CM frame. $\vec{p}$, $\vec{p}_+$, and $\vec{p}_-$ are the proton--$\lis$ relative momentum and the momenta of $e^+$ and $e^-$. Given $|\vec{p}|$, there is one degrees of freedom (DOF), $\theta$, in the photon production, and four in the pair production: $\theta$, $\tpm$, $\phi$, and positron energy $E_+$ [electron energy $E_-= \omega-E_+$ with $\omega$ being the (virtual) photon's energy]. The total pair production cross section can then be computed through~\cite{Rose:1949}:  
\begin{eqnarray}
\sigma_{\ms e^+e^-} &=& \frac{\mr}{p} \frac{\alpha}{16\pi^3} \int  d E_+  d\cos\tpm\, d\cos\theta\, d\phi \notag \\ 
                        &&  \hspace{2cm} \times \frac{p_+ p_-}{8}\, \sum_\mathrm{spins} |\mathcal{M}_{\ms e^+e^-}|^2 \  . \label{eqn:totalxsection}
\end{eqnarray}
Since the original experimental report \cite{Krasznahorkay:2015iga} shows data vs. $\tpm$ and the pair's invariant mass  $\mpm \equiv \sqrt{\omega^2- \left(\vec{p}_+ + \vec{p}_- \right)^2}$ separately, formula for computing $d\sigma$ vs. $d\mpm$ and $d\tpm$ are needed. To calculate $d\sigma/d\mpm$ based on Eq.~(\ref{eqn:totalxsection}), the relation, $p_+ p_-  d E_+  d\cos\tpm  = q p_+' d \mpm  d\cos\theta_+'$, could be used; the ``primed'' variables are measured in the \ep  CM frame, e.g., $p_+'=p_-'=\sqrt{\mpm^2/4-m_e^2}$ with $m_e$ as the electron mass. In the phase space where $\cos\tpm <0$ and $E_+, E_- \gg m_e$,  $m_e=0$ approximation can be applied to simplify the relationship between $E_+$ and $\mpm$ at fixed $\tpm$: $d E_+/d \mpm   =  \mpm  / [\omega |y| \left(1-\cos\tpm\right)] $ with $y\equiv (E_+-E_-)/\omega$, which is then used to compute $d\sigma/d\mpm d\cos\tpm$ based on Eq.~(\ref{eqn:totalxsection}). The allowed phase space is shown in the bottom panel of \fig{fig:kinematics}: given a negative $\cos\tpm$, $4 m_e^2 \leq \mpm^2 \leq \omega^2 (1-\cos\tpm)/2$. We can see that the large-$\mpm$ events have large $\tpm$, while the large-$\tpm$ events have $\mpm$ from $4 m_e^2$ to its upper bound and part of the Jacobian factor, $\mpm p_+ p_- /|y|$, enhances the contribution from the large $\mpm$ region. Although Ref.~\cite{Krasznahorkay:2015iga} shows that the anomaly exists in the large $\mpm$ ($\tpm$) region of $d\sigma/d\mpm$ ($d\sigma/d\cos\tpm$) distribution, it should be informative to see where the anomaly resides in the joint ($\mpm$, $\tpm$) phase space. Note for a fixed $y$, $\mpm^2= (1-y^2) \omega^2 (1-\cos\tpm)/2$, which corresponds to a straight line in the phase space intersecting the horizontal axis at $\cos\tpm=1$, e.g., the solid curve ($y=0$) in the plotted phase space. 

The key quantity in modeling is the EM current's matrix element, $\langle \mathrm{Be}8; -\vec{q} | \hat{J}^{\mu}(\vec{q})| \mathrm{Li7+p};\, a,\,\sigma,\, \vec{p} \rangle$ with $a$ and $\sigma$ as $\lis$ and proton spin projections and  $\vec{q}$ as the (virtual) photon momentum. The matrix element has different components, denoted as  $U_{\lambda S L}$ with $\lambda$, $S$, and $L$ labeling virtual photon's multipolarity, initial state's total spin and angular momentum. In the $J^\pi$ notation, $\lis$, proton, $\bee$ GS, and its excited states of interest are $\frac{3}{2}^-$, $\frac{1}{2}^+$, $0^+$, and two $1^+$s \cite{Tilley:2002vg,Tilley:2004zz}. As dictated by the parity conservation and Wigner-Eckart theorem, the E1 transition is between the s-wave ($L=0$) proton--$\lis$ scattering state and the $\bee$ GS (d-wave should be small), and the total spin $S$ can only be $1$; for the M1 transition $L=1$ and $S=1$ or $2$. The role of E2 transition is also explored here, whose $L=1$ and $S=1$ or $2$. In total, five amplitudes need to be addressed, $\usl{10}$ for E1, $\usl{11}$ and $\usl{21}$ for M1, $\uslt{11}$ and $\uslt{21}$ for E2. 

It is worthwhile to mention a few momentum (length) scales in the reactions. The $\lis$ GS is $2.467$ MeV below its breakup threshold---to ${}^{4}\mathrm{He}+{}^{3}\mathrm{H}$ \cite{Tilley:2002vg}---which translates to a binding momentum $\Lambda \approx 10^2$ MeV if $\lis$ is considered as the bound state of the fragments; the corresponding length scale is $2$ fm. Meanwhile, the $\bee$'s {\it mostly} iso-scalar (MIS) and iso-vector (MIV) $1^+$ resonances are $E_{\ms \left(0\right)}=0.895$ and $E_{\ms \left(1\right)}=0.385$ MeV above the proton--$\lis$ threshold \cite{Tilley:2004zz}; the associated momenta $p$ are about $40$ and $25$ MeV (5 and 8 fm in length scale). By treating $\Lambda$ as the high momentum scale, the $1^+$ states can be considered as composed of ``point'' particle $\lis$ and proton  in the EFT framework.  The $\bee$ GS is $E_{th}=17.2551$ MeV below the proton--$\lis$ threshold and dominated by two ${}^{4}\mathrm{He}$ cluster configuration \cite{Pastore:2014oda,Wiringa:2013fia}, and thus a deep bound state in terms of the proton--$\lis$ configuration. Therefore, the transitions between the $1^+$ states and the $\bee$ GS happen in short distance as compared to 5 fm. These observations suggest that the reactions can be studied in the EFT framework, in which fields with the corresponding parity and spin are assigned to the involved particles and used to construct interaction operators in the lagrangian satisfying rotational, Galilean, parity, and time reversal invariance. (This approach has been successfully applied to study $^8\mathrm{Li}$ and $^8\mathrm{B}$ systems \cite{Zhang:2015ajn}.)  It should be pointed out that near the proton--$\lis$ threshold, the Coulomb interaction between the $\lis$ and proton in the incoming channels needs the standard nonperturbative treatment, i.e., using the Coulomb wave function instead of the plane wave in the Feynman diagram evaluation \cite{Zhang:2015ajn}.  

The relevant Lagrangian is collected here: 
\begin{widetext}
\begin{eqnarray}
\mathcal{L}_{\ms 0}&=&\fdu{n}{\sigma} \left(i\partial_{t}+\frac{\bigtriangledown^{2}}{2\mn} \right) \fd{n}{\sigma} + \fdu{c}{a}\left(i\partial_{t}+\frac{\bigtriangledown^{2}}{2\mc} \right) \fd{c}{a}+ \phi^\dagger \left(i\partial_{t}+\frac{\bigtriangledown^{2}}{2\mn}+E_{th} \right) \phi  +  \psi_{\ms \left(0\right)}^{\dagger\,i}\left(i\partial_{t}+\frac{\bigtriangledown^{2}}{2\mnc} - \Delta_{\ms \left(0\right)} \right) \psi_{{\ms \left(0\right)}\,i} \notag \\
&&+ \psi_{\ms \left(1\right)}^{\dagger\, i}\left(i\partial_{t}+\frac{\bigtriangledown^{2}}{2\mnc} - \Delta_{\ms \left(1\right)} \right) \psi_{{\ms \left(1\right)}i} \ ,\label{eqn:L0} \\ 
\mathcal{L}_{\ms P}&=& \h{0}{3} \psi_{\ms \left(0\right)}^{\dagger\,i} \Tdagger{k j}{i} \Tdagger{a \sigma}{k} \fd{c}{a} V_j \fd{n}{\sigma} + \h{0}{5} \psi_{\ms \left(0\right)}^{\dagger\,i} \Tdagger{\alpha j}{i} \Tdagger{a \sigma}{\alpha} \fd{c}{a} V_j \fd{n}{\sigma}   + \h{1}{3} \psi_{\ms \left(1\right)}^{\dagger\,i} \Tdagger{k j}{i} \Tdagger{a \sigma}{k} \fd{c}{a} V_j \fd{n}{\sigma} + \h{1}{5} \psi_{\ms \left(1\right)}^{\dagger\,i} \Tdagger{\alpha j}{i} \Tdagger{a \sigma}{\alpha} \fd{c}{a} V_j \fd{n}{\sigma} \ ,  \label{eqn:LP}  \\ 
\mathcal{L}_{\ms M1} &= & \dm{0}  \phi^\dagger \mathcal{B}^i \psi_{{\ms \left(0\right)}i} + \dm{1}  \phi^\dagger \mathcal{B}^i \psi_{{\ms \left(1\right)}i} \ ,  \label{eqn:LM}   \\ 
\mathcal{L}_{\ms E1} &=& -i \deone \phi^\dagger  \mathcal{E}^i\, \Tdagger{a\,\sigma}{i} \fd{c}{a} \fd{n}{\sigma} +i \frac{\deone'}{V_\Lambda^2} \phi^\dagger  \mathcal{E}^i\, \Tdagger{a\,\sigma}{i} \fd{c}{a} \vec{V}^2  \fd{n}{\sigma}   \ ,  \\ 
\mathcal{L}_{\ms E2} & = & \detwo{1}  \phi^\dagger  \left(\partial^j \mathcal{E}^i\right)\, \T{\alpha}{ij} \Tdagger{lk}{\alpha} \Tdagger{a\sigma}{l} \fd{c}{a} V_k \fd{n}{\sigma}  + \detwo{2} \phi^\dagger \left( \partial^j \mathcal{E}^i \right) \T{\alpha}{ij} \Tdagger{\beta k}{\alpha} \Tdagger{a\sigma}{\beta} \fd{c}{a} V_k \fd{n}{\sigma}  \ .  
\end{eqnarray}
\end{widetext}
The complex conjugation of the interaction terms are not explicitly shown. In the fields, $n_\sigma$ (proton), $c_a$ ($\lis$), $\phi$ ($\bee$ GS), $\psi_{{\ms \left(0\right)} i}$ (the MIS $1^+$ resonance), $\psi_{{\ms \left(1\right)} i}$ (the MIV $1^+$ resonance), the indices are the particle spin projections;  ``$i,j,k,l,m,n$'' are reserved for $J=1$ multiplet, ``$\alpha,\beta$'' for $J=2$, ``$a,b,c$''  for $J=\frac{3}{2}$, and ``$\sigma,\delta$''  for $J=\frac{1}{2}$.  $\T{\cdots}{\cdots}$ are the C-G coefficients, e.g., $\T{i}{j\alpha} \equiv \langle 1\,j,2\,\alpha |1\, i\rangle$ and $\Tdagger{j\alpha}{i} \equiv \langle 1\, i |  1\,j,2\,\alpha \rangle$. The repeated indices indicate contraction; the resulting scalar is invariant under rotation. Also note the space time metric, $g_{\mu\nu}=diag(1,-1,-1,-1)_{\mu\nu}$ with $\mu,\nu=t,x,y,z$; specifically for the time index, $t$ is used for the sub(super)script, instead of $0$, which is reserved for the spin projection index with $m_j=0$. 

$\mathcal{L}_{\ms 0}$ is the free Lagrangian. The zero energy reference is the proton--$\lis$ threshold. $M_n$ and $M_c$ are the masses of proton and $\lis$, and $M_{nc}\equiv M_n+M_c$; $\Delta_{\ms \left(0,1\right)}$ are the bare self-energies of the $1^+$ fields. $\mathcal{L}_{\ms P}$ includes the p-wave interactions between proton--$\lis$ and the resonances, which can be used to compute the proton--$\lis$ scattering T-matrix in the $1^+$ channels with $S=1$ and $2$. The inelastic channel involving the $\lis$'s first excited state (0.4776 MeV above its GS) is not included here, considering the previous phase shift analysis \cite{Brown:1973xfg} didn't find significant in-elasticity for the energy between 0.4 to 2.5 MeV. In the interaction terms, $\vec{V}$ is the relative velocity operator\footnote{For a general (complex) vector, $\vec{u}\equiv u^x \vec{e}_x + u^y \vec{e}_y + u^z \vec{e}_z$, we can define:  $u^{+1} \equiv  -{(u^x+i u^y)}/{\sqrt{2}}$, $u^{0} \equiv  u^z  $, and  $u^{-1} \equiv  {(u^x-i u^y)}/{\sqrt{2}}$, and introduce a metric, $ \delta^{ij}=\delta_{ij} \equiv (-1)^i \hat{\delta}_{i,-j} $ with $\hat{\delta}$ as the Kronecker delta to raise and lower the rank-1 indices. This metric differs from the one in Ref.~\cite{Edmonds1974} by an extra -1 factor. The antisymmetric tensor, $\epsilon^{ijk}= -i \,\delta_{\mathcal{P}}$ with $ijk=\mathcal{P}(+1,0,-1)$, which also equals $-\sqrt{2}\, i\, \T{i}{jk}$; $\left(\vec{u}\times\vec{v}\right)^i = \epsilon^{ijk} u_j v_k$. }, $ \vec{V} \equiv - i \left(\overrightarrow{\vec{\partial}}/\mn- \overleftarrow{\vec{\partial}}/\mc \right)$. The coupling strengths are constrained by the resonances' strong decay widths through e.g.,    
$ \Gamma_{\ms \left(0\right)}  =  3 \left(C_{\eta_{\left(0\right)},1}\right)^2 p_{\ms \left(0\right)}^3 \left[\left(\h{0}{3} \right)^2 +\left(\h{0}{5} \right)^2\right]/\left(\pi\mr\right) $ for the MIS resonance. Here $\mr \equiv M_n M_c/M_{nc}$, $\eta_{\left(0\right)} \equiv k_c/\omega_{\ms\left(0\right)}$ with $k_c\equiv Z_\mathrm{Li} Z_{p}\alpha_{em} \mr$ and $ \omega_{{\ms \left(0\right)}} \equiv E_{th} + E_{\ms \left(0\right)} $; $p_{\ms \left(0\right)} \equiv \sqrt{2\mr E_{\ms \left(0\right)}}$; $C_{\eta,l}= 2^{l}e^{-\frac{\pi}{2}\eta}|\Gamma(l+1+i\eta)|/\Gamma(2l+2)$ as related to the Coulomb barrier penetrability. Similar formula connects the MIV resonance width to $\h{1}{3}$ and $\h{1}{5}$. 
 $\mathcal{L}_{\ms M1}$ collects the M1 transition vertices between the $1^+$ resonances and the GS, with  $\mathcal{B}_i$ as magnetic field and defined as $\left(\vec{\partial} \times \vec{\mathcal{A}}\right)_i$. The couplings $\dm{0,1}$ can be fixed by the resonances' radiative decay widths through  $\Gamma_{\gamma {\ms \left(0\right)}} =  \dm{0}^{\, 2}  \omega_{{\ms \left(0\right)}}^3/\left(3\pi\right) $ and similar formula for the MIV resonance.  The values of the strong and EM decay widths can be found in Table~\ref{tab:respara}.
Inside $\mathcal{L}_{\ms E1}$ and $\mathcal{L}_{\ms E2}$ are the vertices for the E1 and E2 transitions. $\mathcal{E}_i$  is the electric field, defined as $ \partial_t \mathcal{A}_i -\partial_i \mathcal{A}_t	 $. The extra factor ``$i$'' in $\mathcal{L}_{\ms E1}$ is the result of the time reversal invariance. In $\mathcal{L}_{\ms E2}$, the $\vec{V}$  inserted between $c$ and $n$ requires $L=1$ in the initial state, while the gradient operator applied to the electric field constructs a rank-2  operator.  As discussed before, the two transitions occur at short distance, and thus the contact interaction terms are proper choices. The  $d'_{E1}$ coupling introduces a linear dependence on the CM energy to the E1 amplitude on the top of the constant $\deone$ contribution; the resulted energy dependence of the E1 photon production cross section is consistent with the data (see Fig.~\ref{fig:SfacOnShellPhoton}). Note $V_\Lambda \equiv \Lambda/\mr $ with $\Lambda$ as the high momentum scale and chosen to be $\Lambda=100$ MeV hereafter. 

\begin{table}
 \centering
   \begin{tabular}{|c|c|c|c|} \hline
           & $E_{\ms \left(i\right)}$ (MeV) & $\Gamma_{\gamma {\ms \left(i\right)}}$  (eV) & $\Gamma_{\ms \left(i\right)}$ (keV)  \\  \hline
     $i=0$ & 0.895             & $1.9 (\pm 0.4) $              & $ 138(\pm 6) $          \\   \hline
		 $i=1$ & 0.385             & $ 15.0 (\pm 1.8) $            & $10.7 (\pm 0.6)$        \\   \hline
   \end{tabular}
   \caption{The excitation energies and strong and EM decay widths of the $\bee$'s two $1^+$ resonances \cite{Tilley:2004zz}.} \label{tab:respara}
\end{table}

\section{Calibration against the photon production measurements} \label{sec:calibration}

\begin{figure}
	\centering
	 \includegraphics[width=3.5cm]{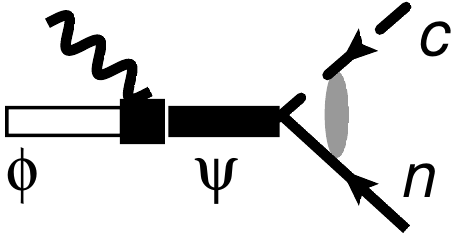}
		\includegraphics[width=2.6cm]{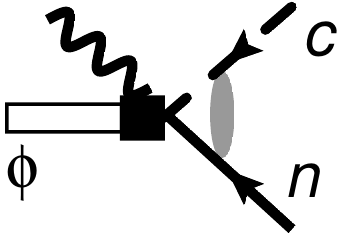}  
     \caption{The Feynman diagrams for the M1, and  E1 and E2 transitions. Here $c$, $n$, $\phi$, and $\psi$ are the fields of $\lis$, proton, $\bee$ GS, and its $1^+$ excited states.} \label{fig:FeyEM}
\end{figure} 

\begin{figure}
		\includegraphics[width=.4\textwidth]{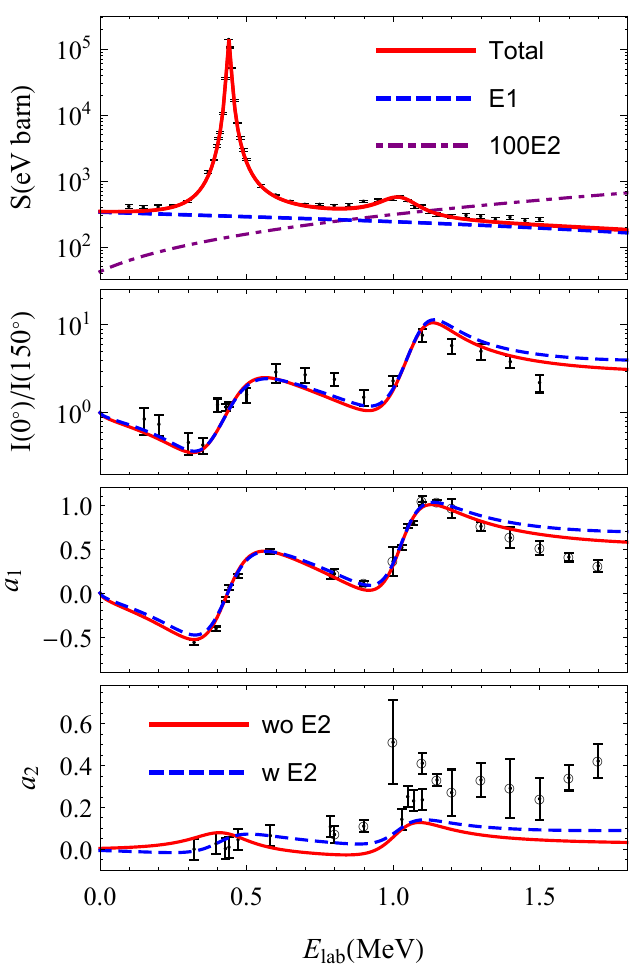} 
     \caption{The $S$ factor (defined in the text) and emission anisotropy of the photon production vs the proton's lab energy $E_\mathrm{lab}$. The data in the top two plots are from Ref.~\cite{Zahnow1995}; in the two lower plots the circled data and the other set are from Refs.~\cite{Schlueter1964} and~\cite{Mainsbridge1960}. } \label{fig:SfacOnShellPhoton}
\end{figure}   	

Fig.~\ref{fig:FeyEM} shows the Feynman diagrams for the M1 (left), and E1 and E2 (right) transitions. The shaded blob means summing diagrams with zero to infinite number of  Coulomb-photon-exchanges, which is equivalent to substituting the plain wave function with  the Coulomb wave function in the diagram evaluation \cite{Zhang:2015ajn}. In the M1-transition diagram, the intermediate $\psi_{\ms \left(0,1\right)}$ propagators, as derived from the Lagrangian $\mathcal{L}_{\ms 0}$ in Eq.~(\ref{eqn:L0}), are simplified with the substitution: $\Delta_{\ms \left(0,1\right)} = E_{\ms \left(0,1\right)} - i {\Gamma_{\ms \left(0,1\right)}}/{2}$.  Summing up these diagrams gives the matrix elements, $\langle \mathrm{Be}8; -\vec{q} | \hat{J}^{\mu}(\vec{q})| \mathrm{Li7+p};\, a,\,\sigma,\, \vec{p} \rangle$, with  
\begin{align}
J^t = & \Tdagger{a\sigma}{i} q^i \usl{10} + q^i  q^j \frac{p_k}{\mr} \T{\alpha}{ij} \big[\Tdagger{lk}{\alpha} \Tdagger{a\sigma}{l} \uslt{11} +\Tdagger{\beta k}{\alpha} \Tdagger{a\sigma}{\beta} \uslt{21}  \big]   \ , \notag \\
J_i  = & \Tdagger{a\sigma}{i} \omega  \usl{10} - \epsilon_i^{\ lm}  q_m\, \frac{p_j}{\mr}  \big[  \Tdagger{\alpha j}{l}  \Tdagger{a\sigma}{\alpha} \usl{21} +  \Tdagger{kj}{l}  \Tdagger{a\sigma}{k} \usl{11}  \big] \notag \\ 
    & + \omega  q^j  \frac{p_k}{\mr}  \T{\alpha}{ij} \big[\Tdagger{lk}{\alpha} \Tdagger{a\sigma}{l} \uslt{11} +\Tdagger{\beta k}{\alpha} \Tdagger{a\sigma}{\beta} \uslt{21}  \big]    \ . \label{eqn:currentME}
\end{align}
In these expressions, 
\begin{align}
\usl{10} &=  \deone \left(1 - d'_{E1} \frac{p^2}{\Lambda^2} \right)\, C_{\eta,0}   e^{i \sigma_0} \ , \\
 \uslt{11}&  =   3 \detwo{1}  \, C_{\eta,1} \,  e^{i \sigma_1} \ , \\ 
 \uslt{21} & =   3 \detwo{2} \, C_{\eta,1} \,  e^{i \sigma_1} \ ,   
\end{align}

\begin{align}
  \frac{\usl{21}}{\sqrt{\mr}} & =   i (3\pi)e^{i \sigma_1} \bigg[ \frac{ C_{\eta,1}}{C_{\eta_{\left(1\right) },1} } \frac{\left[p_{\ms \left(1\right)}\omega_{\ms \left(1\right)}\right]^{-\frac{3}{2}}  \left[\Gamma_{\gamma {\ms \left(1\right)} }  \Gamma_{\ms \left(1\right)} X_{\ms \left(1\right)}\right]^{\frac{1}{2}}}{E-E_{\ms \left(1\right) } + i \frac{\Gamma_{\ms \left(1\right) }}{2} } \notag \\ 
	        & \quad  - \frac{ C_{\eta,1}}{C_{\eta_{\left(0\right)},1} }   \frac{\left[p_{\ms \left(0\right)}\omega_{\ms \left(0\right)}\right]^{-\frac{3}{2}} \left[ \Gamma_{\gamma {\ms \left(0\right)}} \Gamma_{\ms \left(0\right)} X_{\ms \left(0\right)}\right]^{\frac{1}{2}}}{E-E_{\ms \left(0\right)} + i \frac{\Gamma_{\ms \left(0\right)}}{2} }   \bigg]  \ , \label{eqn:resmatrixelement} \\ 
\frac{\usl{11}}{\sqrt{\mr}} & =   i (3\pi) e^{i \sigma_1} \bigg[ \frac{ C_{\eta,1}}{C_{\eta_{\left(1\right) },1} } \frac{\left[p_{\ms \left(1\right)}\omega_{\ms \left(1\right)}\right]^{-\frac{3}{2}}  \left[ \Gamma_{\gamma {\ms \left(1\right)} } \Gamma_{\ms \left(1\right)} \left(1-X_{\ms \left(1\right) }\right) \right]^{\frac{1}{2}} }{E-E_{\ms \left(1\right) } + i \frac{\Gamma_{\ms \left(1\right) }}{2} } \notag \\
                   & \quad +\frac{ C_{\eta,1}}{ C_{\eta_{\left(0\right)},1} }  \frac{\left[p_{\ms \left(0\right)}\omega_{\ms \left(0\right)}\right]^{-\frac{3}{2}} \left[ \Gamma_{\gamma {\ms \left(0\right)} }  \Gamma_{\ms \left(0\right)}   \left(1- X_{\ms \left(0\right) } \right)\right]^{\frac{1}{2}} }{E-E_{\ms \left(0\right)} + i \frac{\Gamma_{\ms \left(0\right)}}{2} }  \bigg]  \ .  \label{eqn:resmatrixelement2}
\end{align}
Note $\omega$ is independent of $|\vec{q}|$ in the above expressions. The notations have been largely explained in the discussion of the Lagrangian, except that (1) $e^{2i\sigma_{l}}\equiv  {\Gamma(l+1+i\eta)}/{\Gamma(l+1-i\eta)} $ is the pure Coulomb scattering phase-shift with angular momentum $l$; (2) $X_{\ms \left(0,1\right) }$ are the branching ratios of the final state $S=2$ channel in the two $1^+$ resonances' strong decay. The relative phases between the two resonances in Eqs.~(\ref{eqn:resmatrixelement}) and~(\ref{eqn:resmatrixelement2}) follow the R-matrix analysis of the photon production in Ref.~\cite{Barker:1996}. The dimensions of the contact couplings are: $\deone \sim [E]^{-5/2}$, and $\detwo{1,2}\sim [E]^{-7/2}$; the EM coupling $e$ is already absorbed into the currents. 

The total squared reaction amplitudes for the photon production, resulted from summing up spin indices and the photon polarization $\tilde{\lambda}$, can be decomposed as 
\begin{eqnarray}
\sum_{a,\sigma,\tilde{\lambda}} |\mathcal{M}_{\ms \gamma}|^2 \equiv T_0 + T_1\, P_1\left(\cos\theta\right) + T_2\, P_2\left(\cos\theta\right) . 
\end{eqnarray}
Here, $P_{n}$ are the Legendre polynomials, and  
\begin{align}
 T_0 &=  2 \omega^2 |\usl{10}|^2 + \frac{2}{3} \omega^2 \left(\frac{p}{\mr}\right)^2 \bigg[|\usl{11}|^2+|\usl{21}|^2 \bigg] \notag  \\
     & \quad + \frac{1}{3} \omega^4 \left(\frac{p}{\mr}\right)^2 \bigg[|\uslt{11}|^2+|\uslt{21}|^2 \bigg]   \ , \label{eqn:T0gamma} \\ 
 T_1 &=  2 \sqrt{2} \omega^2 \left(\frac{p}{\mr}\right)\, \mathrm{Im}\left(\usl{11} \usl{10}^\ast\right) + 2 \omega^3 \left(\frac{p}{\mr}\right)\, \mathrm{Re}\left(\uslt{11}^\ast \usl{10}\right)     \ , \\
 T_2 &=  \frac{1}{3} \omega^2 \left(\frac{p}{\mr}\right)^2 \bigg[|\usl{11}|^2-\frac{1}{5}|\usl{21}|^2 \bigg] \notag \\
     & \quad + \sqrt{2} \omega^3 \left(\frac{p}{\mr}\right)^2 \bigg[ \mathrm{Im}\left(\uslt{11}^\ast \usl{11} \right)-\frac{1}{\sqrt{5}} \mathrm{Im}\left(\uslt{21}^\ast \usl{21} \right) \bigg] \notag \\ 
		 & \quad + \frac{1}{6} \omega^4 \left(\frac{p}{\mr} \right)^2 \bigg[|\uslt{11}|^2-|\uslt{21}|^2 \bigg] \ .  
\end{align}
Note interference exists among E1, M1, and E2, giving rise to $\cos\theta$ modulation and partially to the $P_2$ term. 

The model parameters are fitted against the photon production data, including the total $S$ factor (i.e., $e^{2 \pi \eta}\, E\times$ cross section),  $a_1 \equiv {T_1}/{T_0}$, and $a_2 \equiv {T_2}/{T_0}$ ratios, with the lab energy $E_\mathrm{lab}\equiv {8}/{7} E $ below 1.5 MeV, and with or without E2 transition. We then predict the ratio between the cross sections at $\theta=0^\circ$ and at $150^\circ$. The results are shown in \fig{fig:SfacOnShellPhoton}, while the parameter values can be found in Table~\ref{tab:fittedpara}.  The total $S$ factor and the E1 component in the two fits are indistinguishable in the top panel and shown as the solid red and dashed blue lines. The M1 and E1 in the MIS resonance region have similar contributions, but the former dominates over the latter at the MIV resonance peak. This pattern is consistent with the shell-model \cite{Lawson1980} and the small isospin mixing in the two $1^+$ states \cite{Wiringa:2013fia}. The E2 contribution (purple dotted-dashed line) is on the order of percent of the total at the MIS resonance peak. The slow increase of the E1 component shows  the effect of the $d'_{E1}$ term. In the other three panels, the blue dashed lines are our results with the E2 contribution, while the red solid lines use the other parameter set. We can see the predicted cross section ratios in the second panel and the fitted $a_1$s in the third agree with the data very well, but the two fits underestimate $a_2$ above the MIV resonance in the last panel. Similar observation was also made in Ref.~\cite{Barker:1996}. Including E2 does improves the agreement for $a_2$. However, the two $a_2$ data sets are in minor tension above the MIS resonance, and improved measurements are necessary to determine $a_2$ in this region.

\begin{table}
 \centering
   \begin{tabular}{|c|c|c|c|c|c|} \hline
 $X_{\ms \left(0\right)} $   &  $X_{\ms \left(1\right)} $  & $\deone$  ($\Lambda^{-\frac{5}{2}}$) & $ d'_{E1}$ & $\detwo{1}$ ($\Lambda^{-\frac{7}{2}}$) & $\detwo{2}$ ($\Lambda^{-\frac{7}{2}}$)\\  \hline
            0.574         & 0.712                    & 0.866                                             & 1.64          &  0  & 0 \\   \hline
		        0.629         & 0.743                    & 0.860                                             & 1.61           &  8.27 &  $-18.0$ \\   \hline
   \end{tabular}
   \caption{Two different fits of the model parameters with and without E2 contribution. Here $\Lambda=100$ MeV. } \label{tab:fittedpara}
\end{table}

\section{Electron-positron pair production} \label{sec:epresults}

To study  pair production, we need to attach the lepton line to the photon line in \fig{fig:FeyEM}, i.e., couple the leptonic EM current, $\bar{v}(\vec{p}_+) \gamma^\mu u(\vec{p}_-)$ ($u$ and $v$, and $\gamma^\mu$ are Dirac spinors and matrices), with the nuclear EM current. The total squared reaction amplitude---as the result of summing all the particle's spins---with the photon propagator factorized out, i.e.,  $ \mpm^4 \sum |\mathcal{M}_{\ms e^+e^-}|^2 /2$ is 
\begin{eqnarray}
\sum_{a,\sigma}  \mpm^2\left[\vec{J} \cdot \vec{J}^\ast - J^t J^{t\ast}\right]  - |\Delta^t J^t- \vec{\Delta}\cdot \vec{J}|^2  \ . \notag 
\end{eqnarray}  
Here $\vec{J}$ and $J^t$ as the shorthand of the current matrix elements in Eq.~(\ref{eqn:currentME}), and $\Delta^\mu \equiv p_+^\mu- p_-^\mu$. The above expression  can be expanded as 
\begin{align}
T_{0,0}&+T_{0,2} \cos{2\phi} + T_{1,0}\, P_1   + T_{2,0}\, P_2   
+ T_{2,2}\, P_2  \cos{2\phi} \notag \\
 & + T_{3,1}\, \sin{\theta} \cos{\phi} + T_{4,1}\,\sin{2\theta}\cos{\phi} \ ,  \label{eqn:pairdecomp}
\end{align}
with 
\begin{widetext}
\begin{eqnarray}
 T_{0,0} &=& \bigg[ \mathcal{R}_1\, \mpm^4  + 2  \mathcal{R}_2\, \omega^2 \mpm^2  \bigg] \, |\usl{10}|^2 +  \frac{2}{3} \mathcal{R}_2\,  \mpm^2  q^2\left(\frac{p}{\mr}\right)^2 \bigg[|\usl{11}|^2+|\usl{21}|^2 \bigg]  \notag \\ 
&& + \frac{2}{9}  \left[ \mathcal{R}_1\,\mpm^4  + \frac{3}{2} \mathcal{R}_2\, \omega^2 \mpm^2    \right] q^2 \left(\frac{p}{\mr}\right)^2 \bigg[|\uslt{11}|^2+|\uslt{21}|^2 \bigg]   \ ,  \label{eqn:t00} \\ 
 T_{0,2} &=&  - \frac{2}{3}  {\ssq}\left(\frac{p}{\mr}\right)^2 \bigg[|\usl{11}|^2- \frac{1}{5}|\usl{21}|^2 + \frac{\omega^2}{2} \left(|\uslt{11}|^2-|\uslt{21}|^2\right)   + \sqrt{2} \omega  \left(\mathrm{Im}\,\uslt{11} \usl{11}^\ast - \frac{1}{\sqrt{5}} \mathrm{Im}\,\uslt{21} \usl{21}^\ast  \right)  \bigg] \  , \\
T_{1,0} &=& 2 \sqrt{2}\mathcal{R}_2 \, \omega  \mpm^2 q \left(\frac{p}{\mr} \right)\mathrm{Im}\left(\usl{11} \usl{10}^\ast\right)  + \frac{4}{3} \left[\mathcal{R}_1\, \mpm^4 + \frac{3}{2} \mathcal{R}_2 \, \omega^2 \mpm^2 \right]  q \left(\frac{p}{\mr}\right) \mathrm{Re}\,\left(\uslt{11} \usl{10}^\ast\right)  \ , \\ 
T_{2,0} &=& \frac{1}{3} \mathcal{R}_2  \, \mpm^2   q^2\left(\frac{p}{\mr}\right)^2 \bigg[|\usl{11}|^2- \frac{1}{5}|\usl{21}|^2  + \frac{\omega^2}{2} \left(|\uslt{11}|^2-|\uslt{21}|^2\right)  - 3  \sqrt{2} \omega  \left(\mathrm{Im}\,\uslt{11} \usl{11}^\ast - \frac{1}{\sqrt{5}} \mathrm{Im}\,\uslt{21} \usl{21}^\ast  \right) \bigg]  \notag \\ 
&& + \frac{2}{9} \mathcal{R}_1\, \mpm^4    q^2\left(\frac{p}{\mr}\right)^2 \left(|\uslt{11}|^2-|\uslt{21}|^2\right)  \  , \\ 
T_{3,1} &= & 2 \frac{\mathcal{S}_{+-}}{q}\mpm^2 \Delta^t  \left(\frac{p}{\mr}\right) \bigg[\sqrt{2}\, \mathrm{Im}\,\usl{11} \usl{10}^\ast -\frac{\omega}{3}\, \mathrm{Re}\,\uslt{11} \usl{10}^\ast  \bigg]  \ , \\
T_{4,1} &= & -\frac{1}{3}  \omega  \mathcal{S}_{+-} \mpm^2 \Delta^t\left(\frac{p}{\mr}\right)^2  \bigg[|\uslt{11}|^2-|\uslt{21}|^2+  \frac{ 3  \sqrt{2}}{\omega}  \left(\mathrm{Im}\,\uslt{11} \usl{11}^\ast - \frac{1}{\sqrt{5}} \mathrm{Im}\,\uslt{21} \usl{21}^\ast  \right) \bigg]  \ ,   \\  
T_{2,2}&=&-T_{0,2} \ . \ \mathrm{Here}\ \mathcal{S}_{+-}\equiv p_+ p_- \sin\tpm \ ,  \ \mathcal{R}_1 \equiv  1-\frac{{\Delta^t}^2}{q^2} \ , \  \mathrm{and} \  \mathcal{R}_2   \equiv   1- 2 \frac{\ssq}{q^2 \mpm^2} \ . 
\end{eqnarray}
\end{widetext}
It is reassuring to see that for each multipole, E1, E2, or M1, when neglecting the terms with $\theta$ and $\phi$ dependence in expression~(\ref{eqn:pairdecomp}), the ratio between the pair production cross section based on Eq.~(\ref{eqn:t00}) and the photon production cross section using Eq.~(\ref{eqn:T0gamma}) agrees with the results in Ref.~\cite{Rose:1949}. However the latter study only considered the transitions between nuclear bound or resonance states, while here the beam-target scattering state defines a particular direction leading to the $\theta$ and $\phi$ dependencies. Such angular dependencies and multipole interferences have not been thoroughly studied before (a limited study can be found in Ref.~\cite{Goldring:1952}). 

\begin{figure}
		\includegraphics[width=0.5\textwidth]{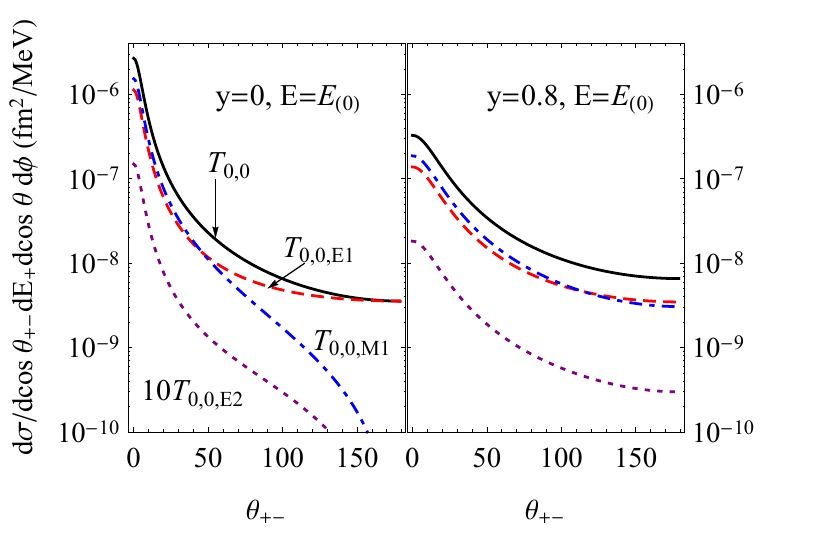}  
     \caption{The $T_{0,0}$'s contribution to the differential cross section vs. $\tpm$ and its decomposition at $y=0, 0.8$ and $E=E_{\ms (0)}$.  $T_{0,0,E2}$ is multiplied by 10 in both plots to increase its visibility.} \label{fig:T00decomp}
\end{figure}

\begin{figure}
				\includegraphics[width=0.5\textwidth]{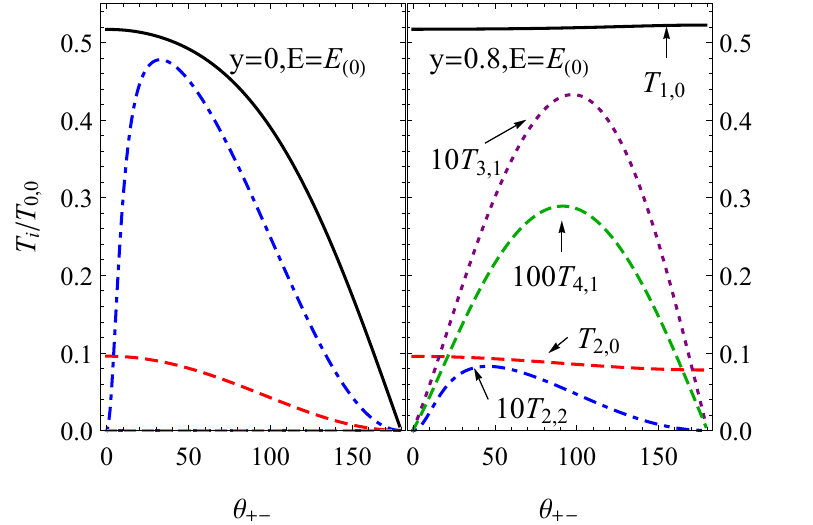} 
     \caption{The ratios between the other coefficients $T_{i,j}$ in expression~(\ref{eqn:pairdecomp}) and $T_{0,0}$ at $y=0, 0.8$ and $E=E_{\ms (0)}$. Several components are multiplied by 10 or 100 in both plots to increase their visibility. Note  $T_{3,1}=T_{4,1}=0$ when $y=0$.} \label{fig:Tidecomp}
\end{figure}

Fig.~\ref{fig:T00decomp}  shows $d\sigma/d\cos\tpm dE_+ d\cos\theta d\phi$ vs $\tpm$ from the $T_{0,0}$ component with $y=0$ and $0.8$ and $E$ fixed at the MIS resonance. The multipoles' contributions are also compared. The E2 contribution is at most a few percent of the total, which is consistent with the photon production results. Although the cross section is dominated by small $y$ and $\tpm$, at large $\tpm$ it is similar at the two $y$ values. Note the upper bound of $\mpm$ for a given $y$ is less than $\omega$ and correspondingly the lower bound of $q$ is above 0, unless $y=0$ (c.f. \fig{fig:kinematics}). Meanwhile, the E1 multipole is $q$ independent and the other two  are proportional to $q$. As the result, the M1 and E2 contributions approach zero towards $\tpm=180^\circ$ (proportional to $1+\cos\tpm$)  when $y=0$, and to a nonzero value when $y\neq 0$. On the other hand, the E1 contribution is always nonzero at  $\tpm=180^\circ$.   Of course, the cross section's $\tpm$ and $y$ dependencies change  with $E$, because the relative weights of the three multipoles depend on $E$; for example at the MIV resonance, the M1 dominates. \fig{fig:Tidecomp} shows the ratio between the terms related to anisotropy and $T_{0,0}$ for the two $y$ values. The dominant term, $T_{1,0}$, gives rise to about $50\%$ modulation, while the $T_{2,0}$ is about $10\%$ or less; the terms associated with $\phi$ dependence are (not surprisingly) proportional to $\mathcal{S}_{\ms +-}$, i.e., the area of the parallelogram spanned by $\vec{p}_{\pm}$, and could reach a few percent of the total in the current chosen kinematics.  The anisotropy is again $E$ dependent, e.g., the $\cos\theta$ modulation turns even bigger above the MIS resonance (could also be inferred from the $a_1$ behavior shown in Fig.~\ref{fig:SfacOnShellPhoton}). Our results show that the pair emission anisotropy is not negligible. It should be included in analyzing the nuclear physics background in this type of experiments looking for new physics. 
Note that  $T_{3,1}$ and $T_{4,1}$ are odd functions of $y$ while all the other components are even functions. Therefore, each term in expression~(\ref{eqn:pairdecomp}) is symmetrical under switching electron and positron and $\phi\rightarrow \phi + \pi$, which is consistent with the charge conjugation symmetry in the single photon approximation.

\begin{figure}
	\centering
		\includegraphics[width=0.5\textwidth]{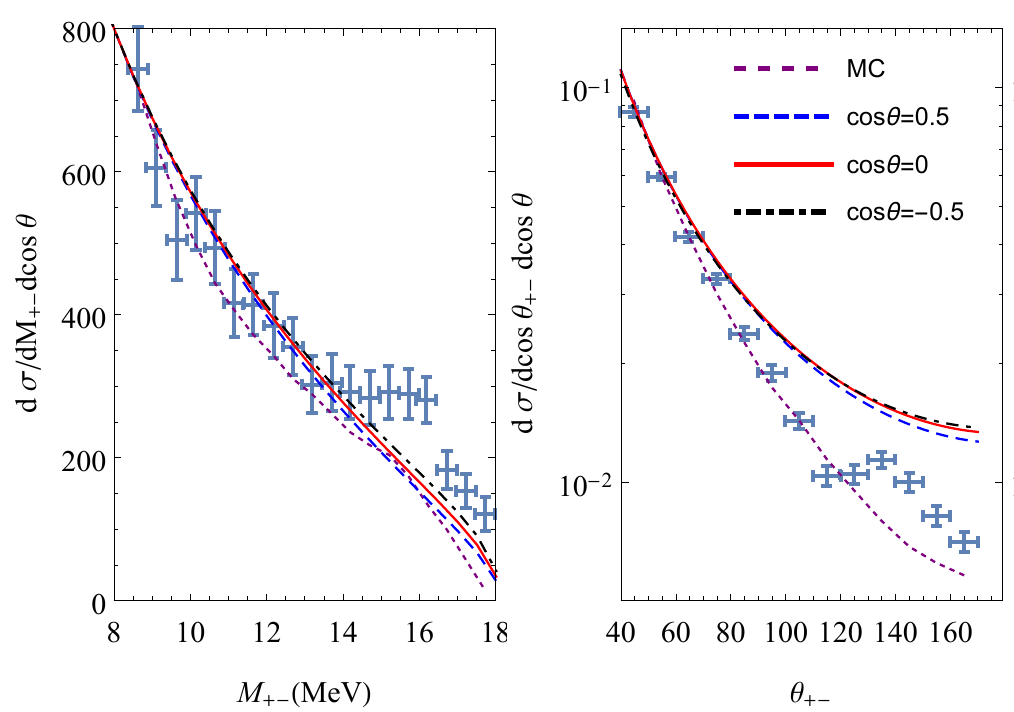} 
     \caption{The differential cross sections vs $\mpm$  (left) and $\tpm$ (right) with $\cos{\theta}=0$ and $\pm 0.5$. ``MC'' is the experimental MC simulation \cite{Krasznahorkay:2015iga}. In the $\mpm$ distribution, the last data point \cite{Krasznahorkay:2015iga} with $\mpm$ above the so-called Q value, i.e., $E_{th}+E_{\ms (0)}=18.15$ MeV, is not shown here. The normalizations of our results in two plots are chosen such that the results agree with data in the lowest $\mpm$ and $\tpm$ bins.} \label{fig:dSdMNoFF}
\end{figure}

Now we can examine whether the $\theta$ modulation (mostly due to the E1-M1 interference), neglected by the experimentalists \cite{Krasznahorkay:2015iga, Gulyas:2015mia}, is related to the observed anomaly. (The $\phi$ modulation can't be addressed in this work.) According to Ref.~\cite{Gulyas:2015mia}, $\theta$ should be $90^\circ$ on average, but  the detector's finite size brings modification. In \fig{fig:dSdMNoFF}, the left panel shows $d \sigma/ d\cos\theta d\mpm$ (with $\phi$ integrated out) with $\cos\theta = \pm 0.5$ and $0$. The normalization of different curves, which can't be fixed without details about the experiment, are adjusted so that they agree with data at $\mpm$ around 8 MeV. We can see decreasing $\cos\theta$ value from 0.5 to $-0.5$ increases the cross section at large $\mpm$ region, but it does not explain the data from Ref.~\cite{Krasznahorkay:2015iga}. The right panel shows a similar comparison for  $d \sigma/ d\cos\theta d\cos\tpm $. The agreement between our  $\mpm$ distribution and the corresponding experimental Monte-Carlo (MC) simulation  (purple dashed curve)  is much better than between the $\tpm$ distributions, suggesting that the detector efficiency variation is significant in the shown $\tpm$ range but not in the shown $\mpm$ range. Furthermore, considering there is a $20\%$ uncertainty for $\Gamma_{\gamma {\ms \left(0\right)}}$ (see Table~\ref{tab:respara}), we also vary it within three times of its uncertainty, and don't see any bump structure consistent with the data in both distributions. We should emphasize that the theory prediction is constrained by the  photon production data, and thus comparison between our calculation and the experimental \ep production data in the kinematics without anomaly should be a valuable cross check.

\section{Add form factor (FF) to explain the anomaly} \label{sec:ff}

\begin{table}
 \centering
   \begin{tabular}{|c|c|c|c|} \hline
           & $f_1$  & $f_2$  & $f_3$  \\  \hline
    FF1   & -3.323            & $-5.759 $    & $ 17.95 $          \\   \hline
	 FF2 & -3.305             & 0            &  0   \\     \hline
    \end{tabular}
   \caption{Two different fits of the FF parameters, $f_1$, $f_2$, and $f_3$, with $\theta$ fixed at $90^\circ$ .} \label{tab:fittedff}
\end{table}

\begin{figure}
	\centering
		\includegraphics[width=0.5\textwidth]{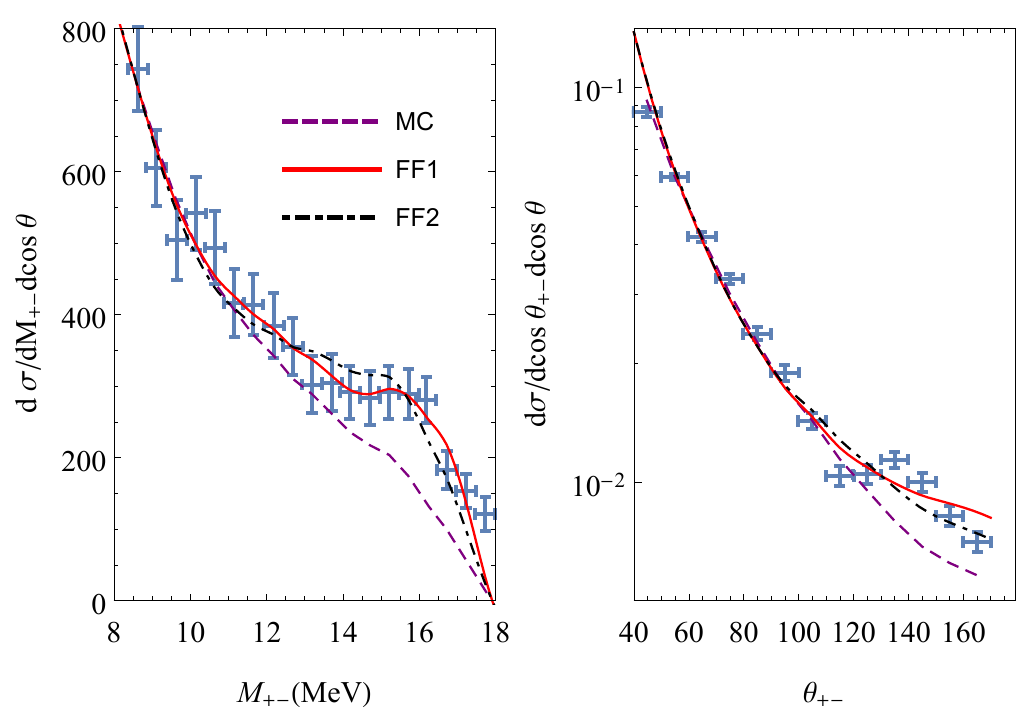} 
     \caption{The differential cross sections vs $\mpm$ and $\tpm$ with $\theta=90^\circ$. Again ``MC'' is the MC simulation. The other curves are explained in the text.} \label{fig:dSdM}
\end{figure}

Here we examine whether introducing a  FF to the resonance's EM coupling vertex $\dm{0}$ might explain the anomaly. (Such FF is not constrained by the photon production measurement.) We utilize a polynomial parametrization, $f(\mpm^2)\equiv 1+ f_1  r  + f_2 r^2 + f_3 r^3 $ with $r\equiv \mpm^2/\tilde{\Lambda}^2$ and $\tilde{\Lambda}=20$ MeV. In order to minimize the impact of missing detector efficiency in our calculation, we fit the ratio, 
\[\frac{d\sigma/d\mpm d\cos\theta\vert_{\theta=90^\circ,\mathrm{with\, FF}}}{\mathcal{N}_1 d \sigma/d\mpm\vert_{\mathrm{without\, FF}}} \quad \mathrm{against}\ \frac{\mathrm{data}}{\mathrm{MC}\ \mathrm{simulation}} \] and extract the FF parameter(s), based on the assumptions that $d \sigma/d\mpm $ calculation without FF should be the closest  to  the MC simulation and that the experimental set up is close to $\theta=90^\circ$. The normalization $\mathcal{N}_1$ is chosen such that the ratio is one at the $\mpm=8.6$ MeV corresponding to the data point with the lowest $\mpm$ value. Two different fits are presented in Table~\ref{tab:fittedff}, both neglecting E2 contribution. In the fitting, the two data points with the highest $\mpm$ values are excluded: the largest-$\mpm$ one has contribution from $\mpm \geq 18.15$ MeV, i.e. above the Q value, and at both data points the ratios between MC and our calculations are dramatically different from those at other data points, indicating significant change of the detector efficiency towards the two largest $\mpm$ bins. We get $\chi^2$ per DOF about 0.5 and 1.1 for the FF1 and FF2 fits. In the left panel of \fig{fig:dSdM}, the curves with different FFs are computed by multiplying the MC simulation with the fitted ratios. In the right panel, the corresponding results for the $\tpm$ distribution are compared. Our results are obtained by multiplying the MC simulation with the ratio \[\frac{d\sigma/d\cos\tpm d\cos\theta\vert_{\theta=90^\circ,\mathrm{with\, FF}}}{\mathcal{N}_2 d \sigma/d\cos\tpm\vert_{\mathrm{without\, FF}}}. \] Note here another independent normalization $\mathcal{N}_2$ is chosen to best match theory with the data. We see that introducing FF allows us to explain the shape of the experiment data binned against $\mpm$ and $\tpm$. However the fitted parameters shown in Table~\ref{tab:fittedff} indicate a momentum scale around $20$ MeV and length scale on the order of $10$s fm, which have not been seen in microscopic calculation, e.g.,Refs.~\cite{Wiringa:2013fia,Pastore:2014oda}. If the natural form factor is used, e.g., $1-\mpm^2/\Lambda^2$, the shape of the calculated $\mpm$ distribution (after renormalization) is changed by less than $1\%$. In addition the absence of the anomaly in the MIV resonance region excludes such length scale for this resonance. 

It should be pointed out that our calculation shows the E1 transition contributes about $50\%$ of the total cross section at the MIS resonance peak, which is much larger than the $23\%$ used in Ref.~\cite{Krasznahorkay:2015iga}. This provides motivation for an attempt to better describe the data by $\, $ simply tuning the normalizations ($\mathcal{N}_{1,2}$) of our results (without introducing FF). The experimental MC simulation results have $\chi^2$ per data-point around 4.3 ($\mpm$) and 16.5
	($\tpm$), while our model can get smaller $\chi^2$ per data-point, 3.3 and 13, by adjusting $\mathcal{N}_{1,2}$.  

In summary, we have improved the previous nuclear physics model for the \ep production in the current experimental context, by including the interferences between E1,  E2, and  M1  multipoles and two different angular dependencies in the modelings, and introducing important constraints from the  photon production measurements. The interferences and emission anisotropy  are currently neglected by the experimental analysis, but could be important for precisely constraining new physics parameters. The approach can also be adapted to study the pair production decaying from new particle and its interplay with the virtual photon decay mechanism, which is also needed in detailed analysis. Moreover, we find that introducing FF to the M1 transition between the MIS resonance and $\bee$ GS is able to explain the shape of the anomaly signal in both $\mpm$ and $\tpm$ distributions, but the length scale associated with the FF is on the order of $10$s fm which has not been seen in the microscopic study of the $\bee$ nucleus. We also notice that tuning the normalizations of our calculations (without FF) reduces the confidence level of the anomaly in both $\mpm$ and $\tpm$  distributions by at least one standard deviation.

\begin{acknowledgments}
We thank Attila Krasznahorkay for clarification of certain details in the original experimental report, and Daniel Phillips, Kenneth Nollett, Carl Brune, Ben Sheff, Yury Kolomensky, Zhaowen Tang, Wick Haxton, Saori Pastore, Robert Wiringa, Martin Savage, and Larry McLerran for discussions and suggestions.  This  work of  was supported by the U. S. Department of Energy Office of Science, Office of Nuclear Physics under Award Number DE-FG02-97ER-41014. 
\end{acknowledgments}


\end{document}